# The Influence of the Aortic Root Geometry on Flow Characteristics of a Bileaflet Mechanical Heart Valve

**Fardin Khalili[*,1], Peshala P.T. Gamage[1], Hansen A. Mansy[1]**

[1]University of Central Florida, Department of Mechanical and Aerospace Engineering, Biomedical Acoustics Research Lab, Orlando, FL, 32816, USA

## ABSTRACT

Bileaflet mechanical heart valves have one of the most successful valve designs for more than 30 years. These valves are often used for aortic valve replacement, where the geometry of the aortic root sinuses may vary due to valvular disease and affect valve performance. Common geometrical sinus changes may be due to valve stenosis and insufficiency. In the current study, the effect of these geometrical changes on the mean flow and velocity fluctuations downstream of the valve and aortic sinuses were investigated. The study focused on the fully-open leaflet position where blood velocities are close to their maximum. Simulation results were validated using previous experimental laser Doppler anemometry (LDA) measurements. Results showed that as the stenosis and insufficiency increased there were more flow separation and increased local mean velocity downstream of the leaflets. In addition, the detected elevated velocity fluctuations were associated with higher Reynolds shear stresses levels, which may increase the chances of blood damage and platelet activation and may lead to increased risk of blood clot formation.

## 1. INTRODUCTION

Many heart disorders initiate within the left ventricle, as this chamber is subjected to the highest mechanical loads [1]. The blood flow through the left ventricle is regulated by the mitral and the aortic valves, which influence the inflow and the outflow conditions, respectively [2]. The aortic valve, in particular, is one of the most commonly affected heart valves in a diseased heart [3]. Aortic valve pathologies such as stenosis and insufficiency cause a variation in the geometry of aortic sinuses and affect the performance of the aortic valve [4]. This incidence is responsible for 44% of morbidity [5]. Analysis of flow dynamics around heart valves [6–9], and cardiac sounds [10–14] may help lowering mortality rates. These pathological changes in extreme cases are often due to aortic incompetence caused by a dilated, aortic dissection, and severe aortic valve stenosis [15–17]. In addition, deformation of the aortic root after valve replacement or structural dysfunction of the recently replaced bioprosthetic heart valve due to pure stenosis due to cusps stiffening is common [18]. It would be desirable to match a prosthetic heart valve type with a specific type of aortic geometry in order to obtain a disturbance-free velocity field with low pressure drop.

Successful analysis of the flow through prosthetic heart valves such as bileaflet mechanical heart valves (BMHVs) depends on sufficient understanding of the conditions under which natural valves function . Previous studies showed that the geometry of the aortic root sinuses can contribute to the vortex generation and flow recirculation [19]. As a result, these shear stresses can cause damage to the blood cells and facilitate thrombus formation. Barannyk et al. [7], analyzed the impact of the aortic root geometries on the pulsatile blood flow through a prosthetic valve. It was found that the different geometries did in fact create different Reynold Shear stresses and recommended that the implantation of a prosthetic valve should be done in conjunction with the root geometry in order to limit the possible levels of stresses.

The objective of this study is to investigate the dimensional changes of the aortic root due to aortic valve disease, such as valve stenosis and valve insufficiency, and to determine the influence of those changes on the appearance of abnormal flow patterns in the flow through aortic bileaflet mechanical heart valve. The accurate representation of a complex aortic root anatomy was modelled as it is essential in order to reproduce the internal physiological flow field correctly. The mean flow and velocity fluctuations downstream of the valve and aortic sinuses with the focus on the fully-open leaflet position were investigated. This information can be used to improve the design of mechanical heart valves in future studies and gain better understanding of the hemodynamics of blood flow through the prosthetic valves.

## 2. MODELS AND METHODS

In this study, a bileaflet mechanical heart valve was modelled similar to previous studies [20–22], which in the fully open position divides the flow into three orifices: two of them (top and bottom orifices) are roughly semicircular and the third (middle orifice) is approximately rectangular (Fig. 1a). An enhancement implemented in the current study (compared to some previous two-dimensional CFD studies [23,24] was to include the valve ring into the model. In addition, a realistic geometry of the aortic sinuses was created since this is important for appropriate flow field analysis [6,25].

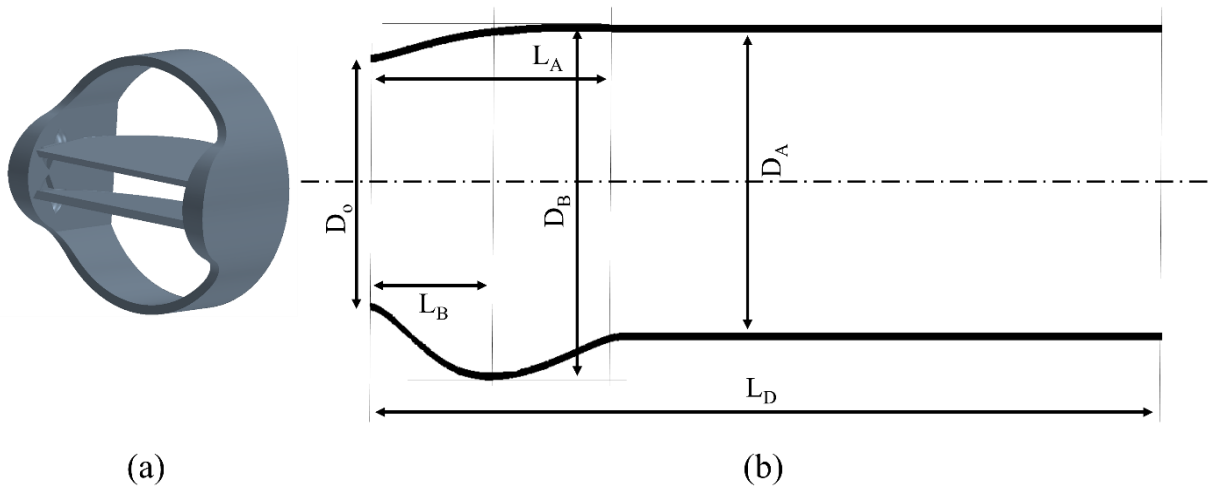

(a)                                                                    (b)

**Fig. 1** (a) Bileaflet mechanical heart valve; (b) Cross-sectional view of aortic root sinuses

Fig.1b shows the cross-sectional view of the asymmetric aortic sinuses geometry with inlet aortic root diameter of d = 0.023 m, which was extracted from angiograms [26]. In this paper, the aortic root was modeled based on following parameters: $D_O$ is the diameter of aortic annulus, $D_A$ is aortic diameter, $D_B$ is the maximum projected sinus diameter, $L_A$ is the length of the sinuses, and LB is the distance between $D_O$ and $D_B$. These parameters can be computed based on the aortic annulus diameter ($D_O$), which is the same as the size of the implanted mechanical heart valve. $L_D$ = 100 mm is the length of the region downstream of the heart valve. The corresponding parameters to the aortic valve stenosis and aortic valve insufficiency are included in the Table 1. They are referred as dilated aortic root and constricted aortic root, respectively [26].

The CFD analysis was performed using a commercial CFD software package (STAR-CCM+, CD-adapco, Siemens, Germany) for a pulsatile flow while the inlet velocity corresponded to cardiac output of 5 L.min-1 and heart rate of 70 bpm with a systolic phase duration of 0.3 s [8,24]. The blood flow parameters such as density, dynamic viscosity, and peak inlet velocity were set to 1080 kg.m-3, μ = 0.0035 Pa.s, and 1.2 ms-1, respectively. This corresponds to a peak inlet Reynolds number ($Re_{peak} = \rho U_{peak} d_{inlet} / \mu$) of 8516 and a Womersley number ($W_o = \frac{d}{2}\sqrt{\frac{\omega\rho}{\mu}} = 26.5$; where, $\omega = \frac{2\pi}{T} = 17.21 \, \text{rad.s}^{-1}$, is the frequency of pulsatile flow and T= 0.866 s is the period. The Wilcox's standard-Reynolds k-Omega turbulence model [27–30] with the inlet turbulent intensity of 5% was used to simulate the flow during a complete cardiac cycle. The current study focused on the fully opening period of the leaflets from 60 to 250 ms [31]. The unsteady simulation was performed with a time step of 0.5 ms and 25 iterations per time step to reach the numerical solution convergence of $< \sim 10^{-4}$. In addition, the uncertainty and error calculations in the current simulations was performed based on ASME recommendations [32]. For the discretization error analysis of the maximum velocity in the entire field, the fine-grid convergence index ($GCI_{fine}$) of 0.139% and maximum discretization uncertainty of approximately 7% (in the area close to the leaflets) were observed. These numerical uncertainties are comparable to previous studies [24]. In this study, $y+$ was maintained less than 1 close to all walls including leaflet surfaces ($y+= 0.46$ at the peak flow).

**Table 1** Parameters for the geometrical characterization of the aortic root

| Parameters (mm) → | $D_O$ | $D_A$ | $D_B$ | $L_A$ | $L_B$ |
|---|---|---|---|---|---|
| Normal | 22.3 | 27.7 | 34.6 | 22.3 | 7.6 |
| Severe Stenosis | 22.3 | 33.5 | 38.4 | 23.2 | 12.1 |
| Severe Insufficiency | 22.3 | 23.5 | 30.6 | 18.3 | 12.5 |

The normalized velocity profile along a line located 7 mm downstream of the healthy valve (at the peak systole) is shown in Fig. 2 for a normal functioning valve. The validation of the velocity profiles in the current study was obtained by comparing the result with previous studies with similar geometries and flow conditions [23,33]. The root-mean-square (RMS) of the velocity differences between the current and the previous experimental study was 6.58% of the maximum velocity, suggesting agreement with measured values [33].

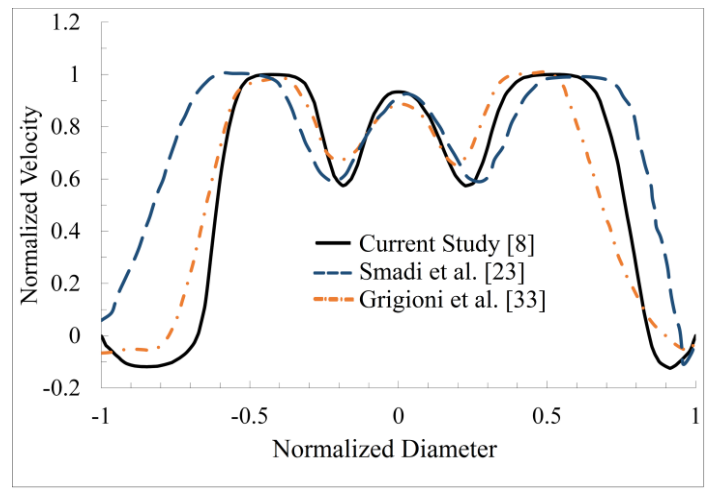

**Fig. 2** Normalized velocity distribution compared to previous CFD [23] and experimental [33] studies.

### 3. RESULTS AND DISCUSSION

Fig. 1 shows the velocity distribution through the BMHV for different aortic root geometries at the peak systole. Results showed a maximum velocity of ~2.5 m.s$^{-1}$ for all geometries which appear at the leading edge of the leaflets and through the three orifices. However, severe stenosis and insufficiency changed the flow pattern downstrem of the heart valve and in the aortic sinuses. For a normal sinuses (Fig. 1a), the flow is relatively uniform downstream of the aortic roots. As the central orifice jet developed during the systole phase, the peak values of velocity fluctuations (or high turbulent intensities) remained concentrated in the wake of the leaflets in the region where the jet became highly unstable and the shear layers breakdown to vortical structures. For the severe insufficiency roots (Fig. 1b), large vortical structures were created and trapped in the sinus region. The high-velocity jets through the top and bottom orifices tend to keep these vortices, and consequently the blood components, inside the sinuses with low velocity and pressure gradients. These may cause higher risk of blood clotting and thrombus formation. In addition, the wake behind the leaflets and high-velocity flow extended far downstream of the leaflets. This can lead to higher wall shear stresses on the aortic sinuses. The velocity distribution downstream of the valve for the severe stenosed aortic sinuses were similar to the normal sinuses (Fig. 1c); however, small-size vortices along with secondary flow region were created. These phenoma can increase the potential for blood damage and platelet activation.

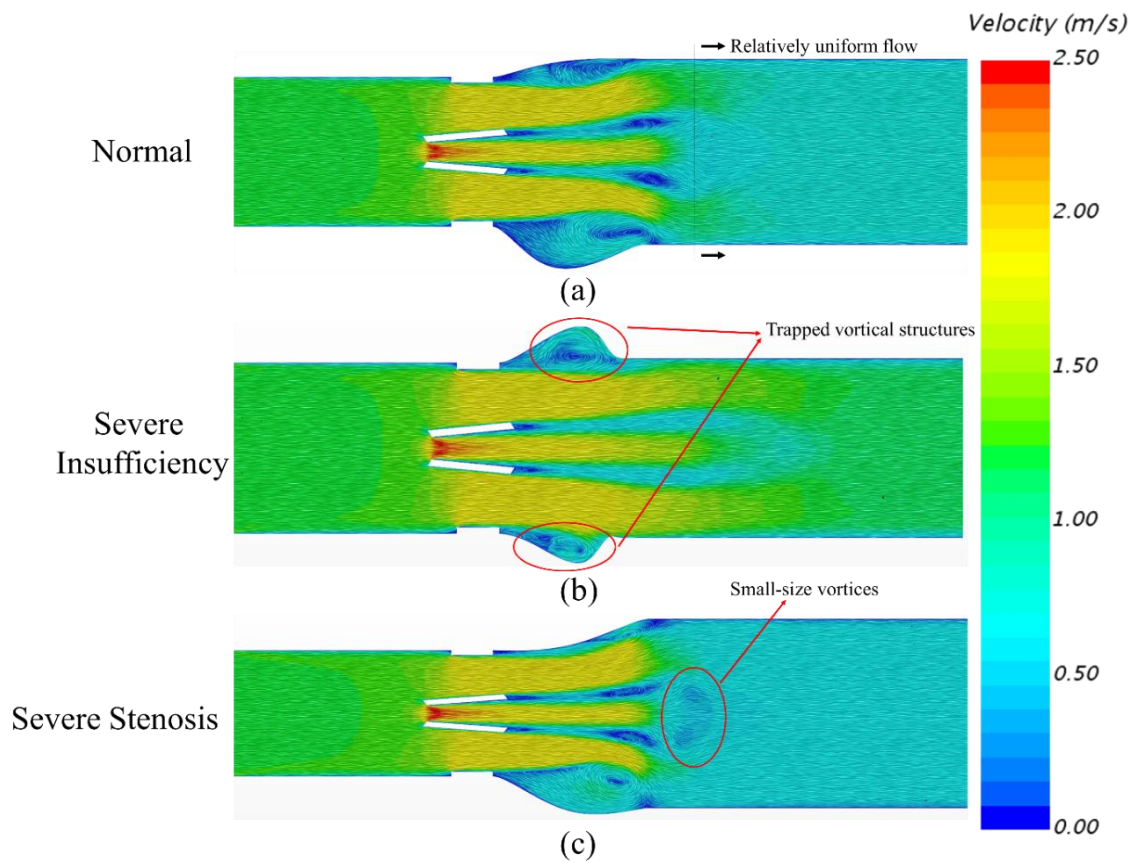

**Fig. 3** Velocity distribution through a bileaflet mehanical heart valve for different geometries of aortic root sinuses: (a) normal sinuses; (b) severe insufficiency; (c) severe stenosis

Fig. 4 shows the instantaneous distributions of wall shear stress (WSS) on the different geometries of aortic sinuses at the peak systolic phase. The vortices which existed in the sinuses (Fig. 3) caused wall shear stresses (WSS) up to 60 Pa for the normal aortic roots (Fig. 4a). The regions with low levels of WSS demonstrate the desirable hemodynamic conditions of this BMHV. For the aortic insufficiency, the trapped vortices (a region of recirculation with high velocity fluctuations) led to high values of WSS about 110 Pa. Conversely, for severe stenosis, wall shear stresses on the sinus were lower and similar to normal sinuses possibly due to lower eddies in the sinuses. High wall shear stresses increase the potential risk of blood clotting and vascular diseases like aortic stenosis.

Several studies reported that the hemolysis (the breakage of a red blood cell membrane), can occur for turbulent shear stresses in the range from 400 to 5000 N.m-2 with exposure time as small as 10 ms [34,35]. In addition, these high turbulent shear stresses can lead to platelets activation, which increase the risk of platelet aggregation and blood clots formation [36]. Clots may detach and the resulting free-floating clot can block arteries leading to serious consequences such as embolism and stroke [37,38]. While stresses acting on the fluid occur in different directions, principal stresses are the highest. Fig. 5 displays maximum turbulent shear (TSS) principal stresses for different geometries of aortic root sinuses at the peak systole and how a deformation in the aortic root geometry led to the elevated levels of the TSS. The TSS distribution through the BMHV for the normal and severe stenosis (Fig. 5a and Fig, 5c) showed similar pattern with the maximum values of ~790 and 805 N.m$^{-2}$. For the severe stenosis, the TSS decreased far downstream of the valve which can indicate the suitability of this BMHV for this condition. On the other hand, TSS is significantly higher around and downstream of the heart valve for the severe insufficiency (Fig. 5b). The maximum TSS value of 820 N.m$^{-2}$ was observed for this aortic root geometry. These results shows that the implantation of this BMHV for the severe insufficiency of the aortic root sinuses needs enough deliberation.

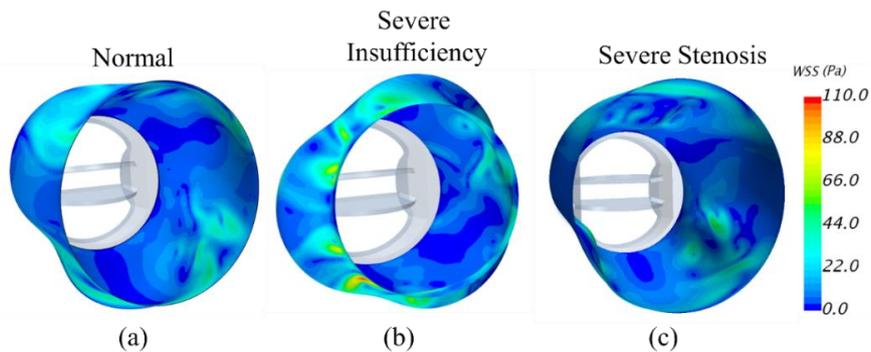

**Fig. 4** Wall shear stress distribution on different geometries of aortic root sinuses: (a) normal sinuses; (b) severe insufficiency; (c) severe stenosis

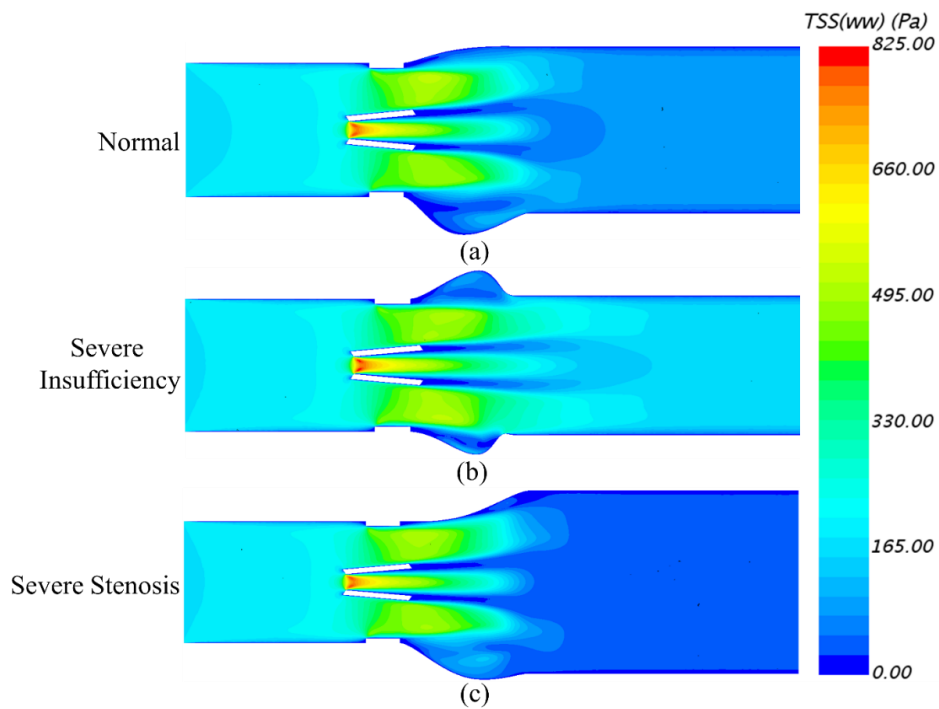

**Fig. 5** Turbulent shear stresses (TSS) through a bileaflet mechanical heart valve for different geometries of aortic root sinuses: (a) normal sinuses; (b) severe insufficiency; (c) severe stenosis

# 6. CONCLUSIONS

In this study, the influence of pathological changes in the dimensions of the aortic root sinuses on the appearance of abnormal flow patterns in the flow through aortic bileaflet mechanical heart valve was investigated. These pathological conditions investigated were valve stenosis and valve insufficiency. The results showed that the flow through the BMHV with normal and aortic root severe stenosis were similar in terms of the vortical structures and corresponding stresses on and downstream of the aortic sinuses. These results demonstrate the desirable hemodynamic conditions of this BMHV for these conditions (normal and severe stenosed aortic roots). On the other hand, the results for the valve insufficiency indicated that flow through the BMHV lead to trapped vortical structures in the sinus region while the turbulent intensity remains high downstream of the valve. Therefore, implanting a heart valve without considering the consequences such as adverse hemodynamic conditions in the aortic root geometry caused by valve diseases might result in sublethal or lethal damage to blood components as well as increased risk of platelet activation.